\documentclass[a4paper,10pt]{article}
\usepackage[utf8x]{inputenc}

\usepackage{amsmath,amssymb,bm}
\usepackage{subcaption}

\usepackage{authblk}

\usepackage{graphicx}
\usepackage{float}
\usepackage[left=2.20cm,top=2.20cm,right=2.20cm,bottom=2.20cm,nohead]{geometry}

\title{\textbf{Multi-domain electromagnetic absorption of triangular quantum rings}}

 \author[1,2]{Anna Sitek}
 \author[3]{Gunnar Thorgilsson} 
 \author[1]{Vidar Gudmundsson}
 \author[3]{Andrei Manolescu}

 \affil[1]{\small{Science Institute, University of Iceland, Dunhaga 3, 
              IS-107 Reykjavik, Iceland} }
 \affil[2]{\small{Department of Theoretical Physics, 
           Faculty of Fundamental Problems of Technology,
           Wroclaw University of Technology, 
           50-370 Wroclaw, Poland} }
 \affil[3]{\small{School of Science and Engineering, Reykjavik University, 
           Menntavegur 1, IS-101 Reykjavik, Iceland} }

\usepackage{bm}
\usepackage[dvipsnames]{xcolor}
\newcommand{\blue}{\textcolor{black}}    
\newcommand{\red}{\textcolor{black}}    

\begin{document}

\date{}              
   
\maketitle

\begin{abstract}
\blue{We present a theoretical study of the unielectronic energy spectra,
electron localization, and optical absorption of triangular core-shell
quantum rings.  We show how these properties depend on geometric
details of the triangle, such as side thickness or corners' symmetry.
For equilateral triangles, the lowest six energy states (including spin)
are grouped in an energy shell, are localized only around corner areas,
and are separated by a large energy gap from the states with higher
energy which are localized on the sides of the triangle.  The energy
levels strongly depend on the aspect ratio of the triangle sides,
i.e., thickness/length ratio, in such a way that the energy differences
are not monotonous functions of this ratio.  In particular, the energy gap
between the group of states localized in corners and the states localized
on the sides strongly decreases with increasing the side thickness,
and then slightly increases for thicker samples.  With increasing the
thickness the low-energy shell remains distinct but the spatial
distribution of these states spreads.  The behavior of the energy levels
and localization leads to a thickness dependent absorption spectrum where
one transition may be tuned in the THz domain and a second transition
can be tuned from THz to the infrared range of electromagnetic spectrum.
We show how these features may be further controlled with an external
magnetic field.  In this work the electron-electron Coulomb repulsion
is neglected.}

\end{abstract}

 \section{Introduction}

Polygonal quantum rings are nanoscale structures with diameters
from tens to hundreds of nanometers and lateral thickness down to
few nanometers \cite{Shi15}. Such structures may be achieved due to
the possibility of combining two or more different materials into one
vertical structure, i.e., core-shell nanowires. They contain a
core, itself a quantum wire, which is covered by one or more layers of
different materials. Due to the variety of controllable physical
properties core-shell nanowires have recently turned out to be suitable
building blocks of many quantum nanodevices such as solar cells
\cite{Krogstrup13,Tang11}, nanoantennas \cite{Kim15}, field-effect
transitions \cite{Xiang06,Nguyen14}, lasers \cite{Saxena13}, including
plasmon lasers \cite{Ho15}, light-emitting diodes \cite{Thierry12},
THz radiation sources \cite{Ibanes13} and detectors \cite{Peng15}.

One of the physical properties which may be modeled during the growth
process is band alignment. Core-shell nanowires are built of at least
two different materials, each one having its own energy structure,
which changes considerably in the presence of strain \cite{Chuang95}.
In most cases core and shell materials have different lattice constants,
and thus the combined system differs substantially from the bulk
of each component, or from single material nanowires, and depends on
the geometrical details. This provides a possibility to control band
alignment through the core and shell thicknesses \cite{Pistol08,Wong11}
and thus to achieve systems in which conduction electrons may be found
only in the shell area \cite{Blomers13}. The present art of manufacturing
allows for etching the core part such that the remaining shell forms a
nanotube of finite thickness \cite{Rieger12,Haas13}. If such nanowires
are sufficiently short, i.e., of height much smaller that the radius,
then they may be considered as quantum rings, as long as only the
lowest wave-function mode in the growth direction is relevant.

Physical properties of polygonal quantum rings differ considerably from 
their circular counterparts. Electrons are always equally distributed 
along the whole circumference of homogeneous circular structures, while in
the case of polygonal systems the localization pattern is much more 
complicated. As in the case of bent quantum wires \cite{Sprung}, in the 
corners of polygonal rings effective quantum wells are formed, with energy 
levels determined by the size and shape of the area between internal 
and external polygon boundaries. This results in localization of low energy 
electrons in the vicinity of corners, whereas carriers associated with higher 
energy levels are purely or partly localized in side 
areas \cite{Ballester12,Sitek15}. In the absence of external fields  
the ground state of circular rings is 
twofold (spin) degenerate while all exited states are fourfold 
(spin and orbital momentum) degenerate. In the case of polygonal 
rings the ground state is also twofold degenerate, but higher levels 
are either two- or fourfold degenerate and the degeneracy pattern is 
determined by the number of corners \cite{Estarellas15}. Depending on 
the ring shape, corner-localized states may be separated from 
side-localized ones by an energy gap \cite{Sitek15}. 

In this paper we show how the spectral properties of the single electron 
states confined in triangular quantum rings depend on the geometric aspect 
ratio, i.e., the ratio between the lateral thickness and the side length, or,
equivalently, the radius of a circle encompassing the ring. We show how corner 
localization 
maxima merge with increasing side thickness and define conditions under which 
low energy electrons are expelled from side areas. The ground state energy and 
splittings between consecutive levels depend on the ring's shape. The energy 
gap separating corner- from side-localized states may reach high values for 
thin triangular samples and may be tuned within a wide energy range. 
As a result the rings absorb photons of distinct energies which may be 
controlled to high extent.

The paper is organized as follows. In the next section we define the sample 
model and describe the methods we use in our calculations. In Sec.\ \ref{sec:results}
we show our results, in particular we investigate how energy levels and carrier 
localization change with the aspect ratio between the side thickness and sample radii
for the case of equilateral rings (Sec.\ \ref{sec:energy}) and non-symmetric samples 
(Sec.\ \ref{sec:non}). Next we show how those features are reflected in the absorption 
spectrum, Sec.\ \ref{sec:absorption}. The summary and final remarks are contained 
in Sec.\ \ref{sec:conclusions}.

\section{\label{sec:model} The model}

\begin{figure}\centering
    \includegraphics[width=0.42\textwidth,angle=0]{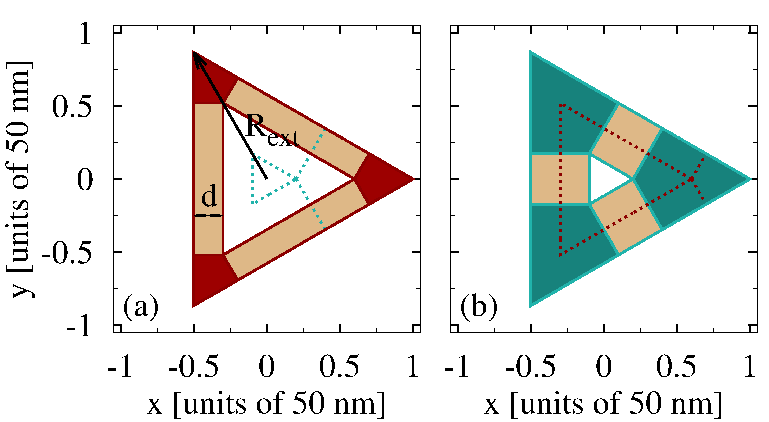}
\caption{ \blue{
          Ring model: Symmetric samples restricted internally and externally 
          by equilateral triangles. We keep the external radii constant and 
          equal $R_{\mathrm{ext}}=50$ nm and change the side thickness $d$. 
          In Fig. (a) $d=10$ nm and in Fig. (b) $d=20$ nm.} 
          The areas 
          included between the internal and external boundaries where 
          effective quantum wells are formed are marked in red or turquoise. 
          Doted turquoise and red lines indicate internal boundaries and 
          corner areas of $20$ (a) and $10$ nm (b) thick samples.
        }   
\label{fig:sample}         
\end{figure}

\blue{ All analyzed rings are restricted externally by equilateral 
triangles, with
a fixed radius $R_{\mathrm{ext}}=50$ nm, i.e., the distance from the
center to the corners, but variable side thickness, such that the core
area is varied, as illustrated in Fig.\ \ref{fig:sample}.  The quantum
states were obtained with two different computational methods, based
on discretization of the Schr\"odinger equation on polar and triangular 
grids, respectively.  }

\blue{In the method based on the polar grid we start with a disk of
radius $R_{\mathrm{ext}}$ discretized
in polar coordinates \cite{Daday11}. The triangular samples are achieved
by imposing polygonal boundaries within the disk, and excluding from
the grid all sites situated outside those boundaries. We have recently
used this method to describe various polygonal rings \cite{Sitek15}.  
In the position representation the Hilbert space is spanned by vectors 
$|kj\sigma\rangle$, where the indexes $k$ and $j$ correspond to discretized radial 
$r=r_k$ and angular $\phi=\phi_j$ coordinates, separated by intervals $\delta r$ and 
$\delta \phi$, and $\sigma$ is the spin projection on the $z$ direction.  
The kinetic Hamiltonian matrix elements defined on the polar grid are 
\begin{equation}
\label{Hamiltonian}
H^{K}_{kj\sigma,k'j'\sigma'} = 
T\delta_{\sigma,\sigma'}\left[ t_r \left(\delta_{k,k'}
-\delta_{k,k'+1}\right)\delta_{j,j'} 
+ t_{\phi}\delta_{k,k'}\left(\delta_{j,j'}-\delta_{j,j'+1}\right) 
+ \mathrm{H.c.}\right], 
\end{equation}
where $T=\hbar^{2}/(2m^{*}R^{2}_{\mathrm{ext}})$ is the reference energy, 
$m^{*}$ the electron mass in the ring material,
$t_r=(R_{\mathrm{ext}}/\delta r)^2$, and
$t_{\phi}=[R_{\mathrm{ext}}/(r_k\delta\phi)]^2$.
We assume that the ring is immersed in an external static magnetic 
field $B$  perpendicular to the plane of the ring, i.e.,
associated with the vector potential $\bm{A}=B(-y,x,0)/2$.  The corresponding 
Hamiltonian matrix elements are
\begin{eqnarray*}
\label{Hamiltonian_B}
H^B_{kj\sigma,k'j'\sigma'} = 
T\delta_{\sigma,\sigma'} \delta_{k,k'} 
\left[\frac{1}{2}t^2_{\mathrm{B}}\left(\frac{r_k}{4R_{\mathrm{ext}}}\right)^2\delta_{j,j'} 
-t_{\mathrm{B}}\frac{i}{4\delta\phi}\delta_{j,j'+1} + \mathrm{H.c.}\right],
\end{eqnarray*}
with $t_{\mathrm{B}}=\hbar e B/ m^{*}T$ being the cyclotron energy in units of $T$.
Finally, we take into account the Zeeman effect which results in splitting of 
spin degenerate energy levels. For the relevant Hamiltonian we obtain
\begin{eqnarray}
\label{Hamiltonian_Z}
H^{\mathrm{Z}}_{kj\sigma,k'j'\sigma'} = 
\frac{1}{2}Tt_{\mathrm{B}}\gamma\left(\sigma_z\right)_{\sigma,\sigma'}\delta_{k,k'}\delta_{j,j'},
\end{eqnarray}
where $\gamma=g^{*}m^{*}/2m_{e}$ is the ratio between the Zeeman gap 
and the cyclotron energy, and $m_e$ is the free electron mass.
The total Hamiltonian is thus $H=H^K+H^B+H^Z$.
We use material parameters of InAs which are the effective mass 
$m^{*}=0.023m_{e}$ and the effective g-factor $g^{*}=-14.9$.  }

In the second method the software package KWANT \cite{Groth14} was
used to construct the Hamiltonian matrix elements of the system. The
system was discretized on an equilateral triangular grid. This scheme
allowed for an equally dense grid over the whole triangular quantum ring,
together with a good representation of the corners. The eigenenergies and
eigenvectors of the resulting Hamiltonian matrix where calculated using a
linear algebra package.  

The numerical calculations on the (nonuniform) polar grid were performed
using around $6900$ grid points and between $2000$-$12000$ in the case
of the KWANT software.  We used both computational methods to explore
the dependence of the energy levels and wave functions on side thickness,
and since both methods predicted very similar behavior we present only
the results obtained with the polar grid.

 \section{\label{sec:results}Results and discussion}
 
Below we show our results on the geometry dependent properties of triangular 
quantum rings. We start our discussion from 
symmetric (equilateral) samples with  constant 
external radii and show how their energy levels and electron localization change 
with increasing side thicknesses. In the next step we analyze the electron 
energies and localization for the case of non-symmetric rings. Finally, we show 
how those features affect the absorption spectrum. 

 \subsection{\label{sec:energy}Energy structure and carrier localization of symmetric rings}
 
\begin{figure}\centering
    \includegraphics[width=0.48\textwidth,angle=0]{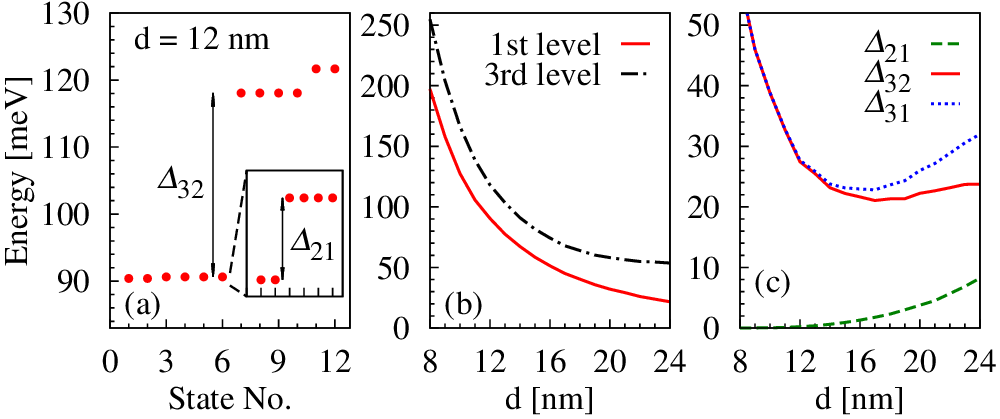}
\caption{ Energy levels of equilateral triangular quantum rings.
          (a) The four lowest energy levels of 12 nm thick sample and their 
          degeneracy, the two lowest levels are repeated in the inset.
          (b) Ground state and the third energy level changing with 
          side thickness. 
          (c) Energy gap dependence on side thickness: energy splitting 
          between the two levels associated with corner-localized 
          probability distributions, $\Delta_{21}$, an energy gap 
          separating corner- from side-localized states, $\Delta_{32}$, 
          and the energy difference between the third energy level and 
          the ground state, $\Delta_{31}$.
        }   
\label{fig:energies}         
\end{figure}

The energy levels of polygonal quantum rings show a unique degeneracy pattern 
which depends only on the number of corners \cite{Sitek15,Estarellas15}. 
In the case of 
symmetric (restricted internally and externally by regular triangles)
rings and zero magnetic field 
the ground state is twofold 
(spin) degenerate, the second and third levels are fourfold (spin and 
orbital momentum) degenerate and the fourth level is again twofold 
degenerate, Fig.\ \ref{fig:energies}(a). The ground state energy and the 
gaps between consecutive levels depend on the aspect ratio of the triangle. 
If the external radius is assumed to be constant then the 
ground state energy changes as the inverse of squared side thickness 
[red solid line in Fig.\ \ref{fig:energies}(b)]. But the evolution of the 
third energy level [the black dashed dotted line in Fig.\ \ref{fig:energies}(b)] is 
more complicated: it decreases relatively fast for thin rings and slows down 
for thicker samples. In Fig.\ \ref{fig:energies}(c) we show the energy intervals 
between the three pairs of levels. The energy gap in the corner state domain,
i.e., between the ground state and the first excited state, $\Delta_{21}$, 
increases with side thickness [the green dashed line in Fig.\ \ref{fig:energies}(c)]. 
The next energy gap, $\Delta_{32}$, which separates corner- from side-localized states, 
initially decreases with the thickness until it reaches a minimum value
for the aspect ratio of around $0.34$, and then it starts increasing slightly
[the red solid line in Fig.\ \ref{fig:energies}(c)].  The sum of those two 
energy gaps makes up a splitting between the ground state and the third energy 
level, $\Delta_{31}$ [the blue dotted line in Fig.\ \ref{fig:energies}(c)], which 
differs considerably from the main energy gap $\Delta_{32}$ only for thick 
samples where the contribution from the splitting of the two lowest levels 
is substantial. In particular the minimum of $\Delta_{31}$ occurs
for a thickness only slightly smaller than  for $\Delta_{32}$.

The mentioned features may be explained with a closer look at the geometry of 
the corner areas.  In Fig.\ \ref{fig:sample} we compare two rings 
which are externally restricted by identical equilateral triangles but have 
different side thicknesses, 10 nm in Fig.\ \ref{fig:sample}(a) and 20 nm in 
Fig.\ \ref{fig:sample}(b), respectively. As can be seen, the corner areas between 
the internal and external boundaries, where effective quantum wells are formed, 
increase with ring thickness. Since quantum well energy levels scale with the 
squared inverse of their size, the decrease of the ground state energy shown 
in Fig.\ \ref{fig:energies}(b) may be understandable.

\begin{figure}[t]\centering
    \includegraphics[width=0.95\textwidth,angle=0]{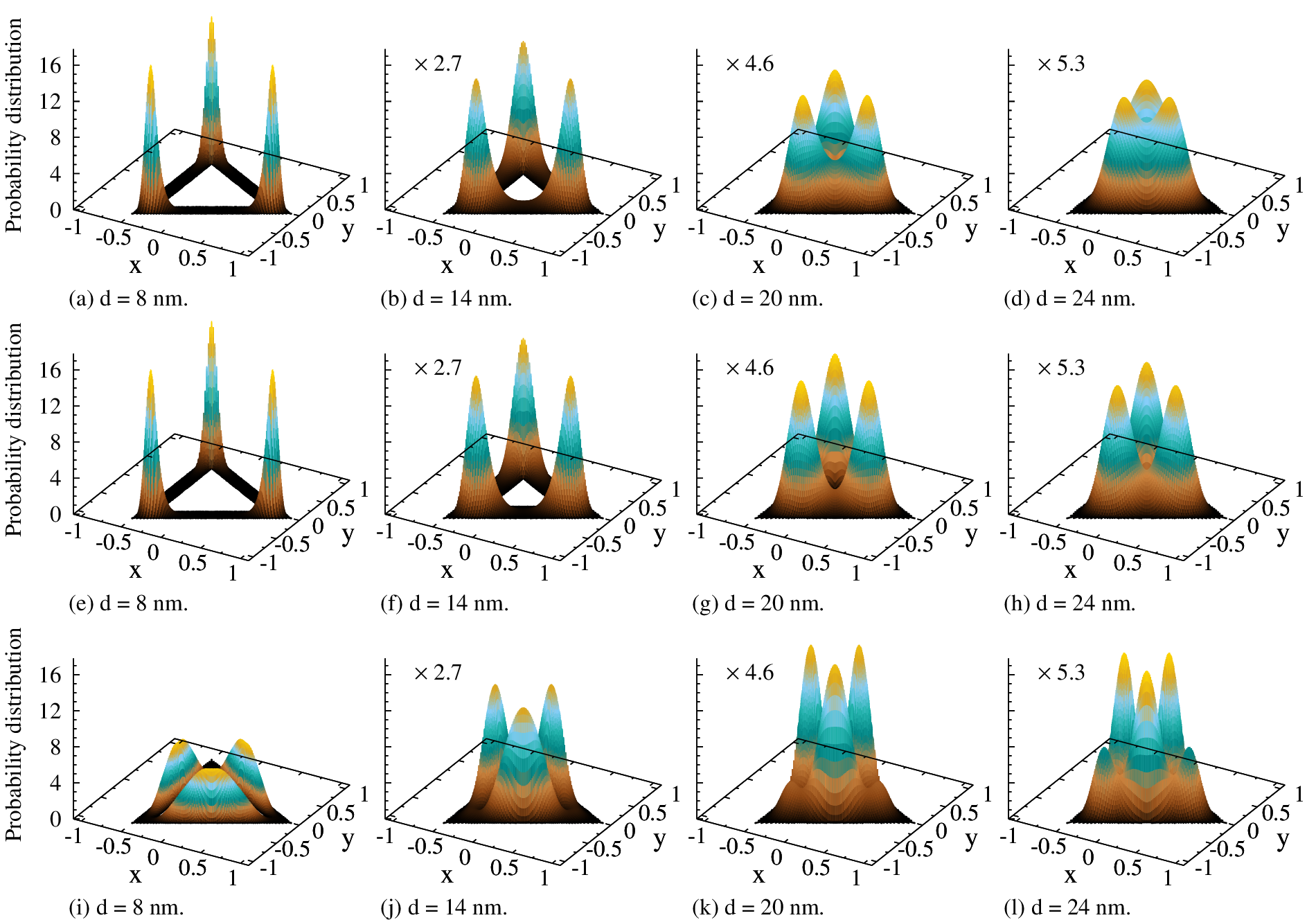}
\caption{\blue{Probability distribution associated with the ground state (a-d), 
         the second energy level (e-h) and the third level (i-l) for
         symmetric triangular rings of different side width $d$.  
         The $x$ and $y$ coordinates are in units of 
         $R_{\mathrm {ext}}$ and probability distribution in units of 
         $R_{\mathrm {ext}}^{-2}$. The distributions for d=8 nm are taken 
         as reference, and for the other cases they are magnified with 
         factors shown on the top left corner of each graph.}
         }    
\label{fig:localization}         
\end{figure}

The localization pattern also changes with increasing side thickness, 
Fig.\ \ref{fig:localization}, but low-energy electrons are always attracted
by the corner quantum wells. The number of corner-localized states equals 
to the double number of corners (if spin degeneracy is included) which means 
that probability distributions associated with the two lowest levels (two- and
fourfold degenerate) of triangular rings form maxima in that 
areas \cite{Sitek15}.  In the case of narrow rings the ground state is 
localized over very small areas in the corners, where sharp and high localization 
peaks are formed, and the probability distribution vanishes along the whole side 
length, Fig.\ \ref{fig:localization}(a).

The corner areas increase with side thickness (Fig.\ \ref{fig:sample})
and the ground state localization maxima decrease, becoming
spread over much larger areas, and penetrating into the sides. For
sufficiently thick samples, i.e., for aspect ratio around $0.24$,
the localization peaks begin merging in the middle of the sides,
Figs.\ \ref{fig:localization}(b)--\ref{fig:localization}(d).
Similar effects occur for the corner-localized probability
distributions associated with the second energy level, Figs.\
\ref{fig:localization}(e)--\ref{fig:localization}(h), but corner maxima
become even sharper with increasing energy, and thus the probability of
finding electrons from the second energy level in the middle of the
sides increases much slower compared to the case of the ground state. The
thicker is the ring the larger is the difference between localization
patterns associated with the two lowest levels which results in higher
energy splitting between the corresponding energy levels [the green 
dashed line in Fig.\ \ref{fig:energies}(c)].

\blue{
According to our definition, thick rings are built around very narrow cores. 
For the 24 nm thick sample shown in 
Figs.\ \ref{fig:localization}(d),\ \ref{fig:localization}(h), and\ \ref{fig:localization}(l) 
the core radius equals 2 nm. The side thickness of 25 nm corresponds to a full 
triangle. The ground state probability distribution for a full triangle forms one, 
relatively wide, peak in the middle of the sample. The localization pattern shown 
in Fig.\ \ref{fig:localization}(d) is split into three half merging maxima 
situated close to the center, i.e.,
looking like the ground state probability distribution of a full triangle pierced 
in the middle. The first excited state of a full triangle, which incorporates shorter
wavelengths, forms sharper maxima shifted towards the corners. The difference of the 
probability distributions associated with the first and second energy levels of 
wider triangular rings can be understood by analogy with the full triangle. 
The result is that the localization peaks in the ground state are (slightly) 
broader than in the first excited state.}

In the case of sufficiently thin samples the probability distribution
associated with the first value above the corner-localized domain, 
i.e., the third energy level, is distributed between all of 
the sides of symmetric samples and forms a maximum in the middle of 
each side \cite{Sitek15}. This localization shape changes with
thickness in the opposite way to the ground state localization pattern, 
Figs.\ \ref{fig:localization}(i)-\ \ref{fig:localization}(l). As seen 
in Fig.\ \ref{fig:sample}, the thicker is the ring the larger is 
the ratio of the corner to side areas 
[red or turquoise to beige areas in Fig.\ \ref{fig:sample}]. 
This results in sharper and higher side maxima which need to adjust to 
relatively smaller localization space. Interestingly,
for thicker rings purely side-localized states do not occur and the 
probability distribution associated with the first level above the 
corner-localized domain forms sharp maxima in the middle of the sides, 
which may be even higher than the peaks associated with the second 
energy level, and are accompanied by lower corner maxima, 
Figs.\ \ref{fig:localization}(k) and \ \ref{fig:localization}(l).

The \blue{most interesting characteristic feature of triangular rings
is the large energy gap separating states with qualitatively different
probability distributions, in the corner areas, or in the middle of the
sides.  Analog corner and side localization occurs for square
and hexagonal rings, but in those cases the energy separating such states is 
much smaller than for triangles \cite{Sitek15}.  }
In the discussion of Fig.\ \ref{fig:localization},
these localization patterns 
change in opposite ways with increasing side
thickness, i.e. the difference between them becomes less pronounced,
which corresponds to the decrease of the energy gap. 
For thick 
rings the probability distribution associated with the third energy level
forms six maxima, i.e., again the localization patterns start to
differ more and thus the energy gap increases slightly.

 \subsection{\label{sec:non}Energy structure and carrier localization of non-equilateral rings}

\begin{figure}\centering
    \includegraphics[width=0.48\textwidth,angle=0]{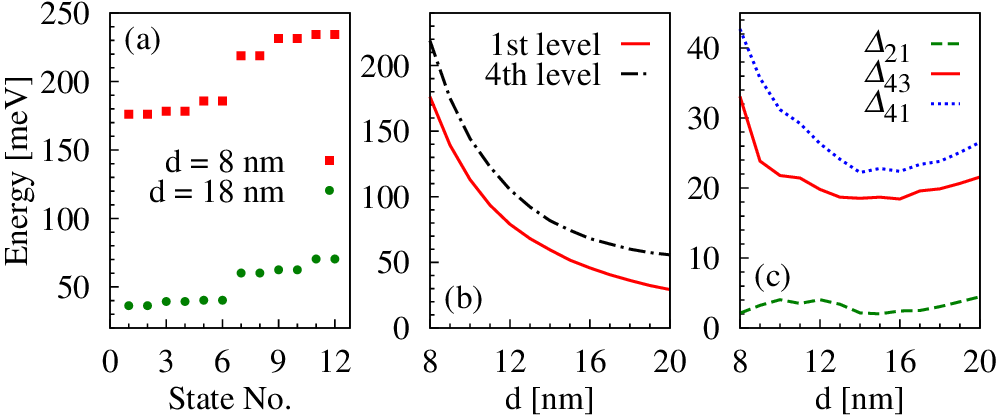}
\caption{ \blue{Energy levels of non-symmetric triangular rings.} 
         (a) The six lowest energy levels  and their degeneracy for two 
         samples with aspect ratios equal to $0.16$ \blue{($d=8$ nm)} and $0.36$ \blue{($d=18$ nm)}, respectively.
         (b) Ground state and the fourth level changing with side 
         thickness. 
         (c) Energy gap dependence on side thickness: energy splitting 
         between the two lowest levels associated with corner localized 
         probability distributions, $\Delta_{21}$, an energy gap separating 
         corner- and side-localized  states, $\Delta_{43}$, and the energy 
         difference between the fourth (mostly side-localized \blue{level}) and 
         the ground state, $\Delta_{41}$.
\red{    The values of the side thickness shown in Fig. (a) and on 
         the axises of Figs. (b) and (c) refer to the
         thickness of the narrowest side.}
         }   
\label{fig:energies_non}         
\end{figure}
\begin{figure}[t]\centering
    \includegraphics[width=0.95\textwidth,angle=0]{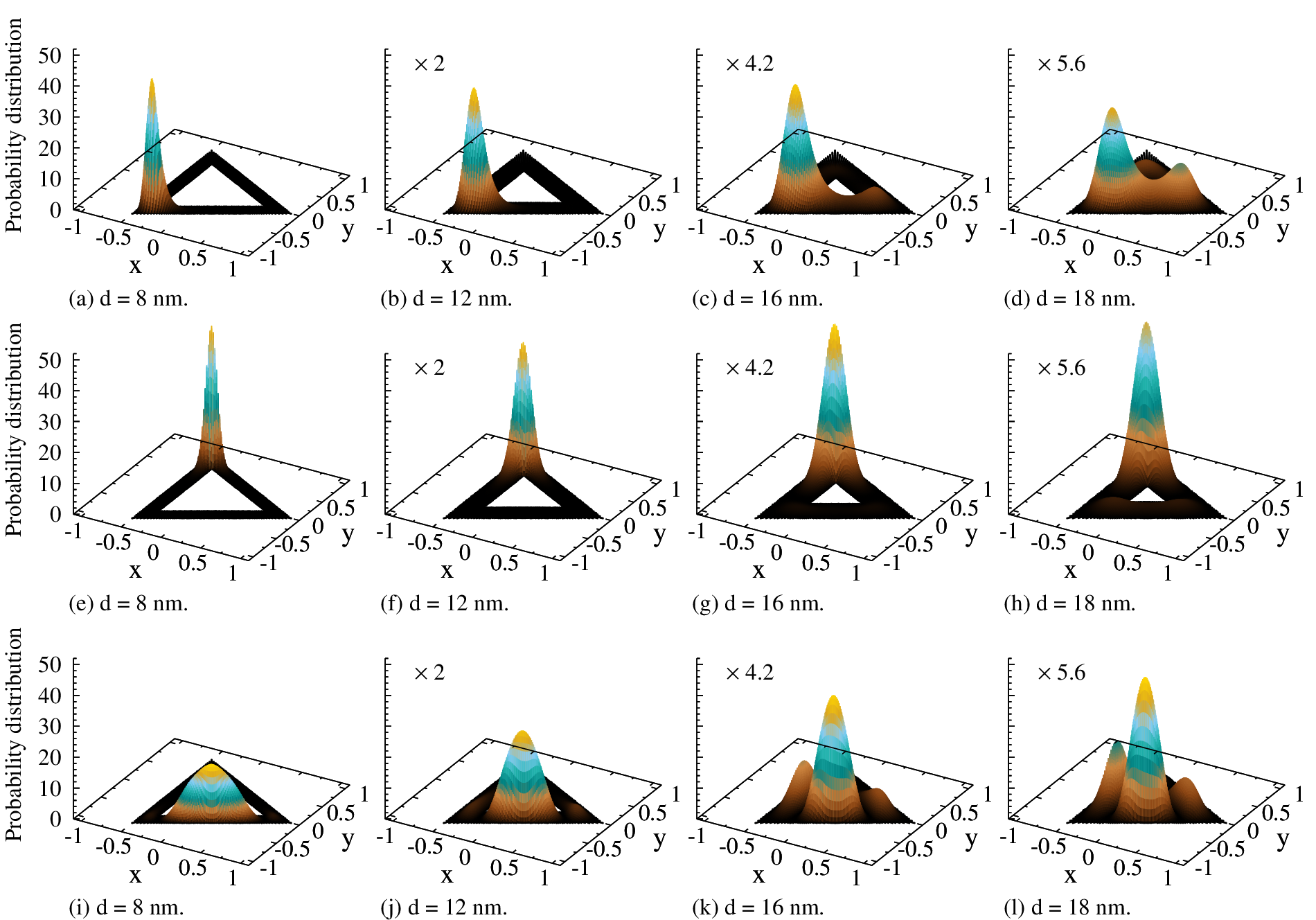}
\caption{Probability distribution associated with the ground state (a-d), 
         the third energy level (e-h) and the fourth one (i-l) for 
         non-symmetric samples.
\red{    The samples were 
         restricted externally by the same          
         equilateral triangles for which $R_{\mathrm{ext}}=50$ nm         
         while the internal boundaries 
         were defined by different side thicknesses. The values of the 
         side thickness shown in the figure captions refer to the sides parallel 
         to lines $y=(-x+1)/\sqrt{3}$, the thicknesses of the two other 
         sides were increased by $4\%$ (sides parallel to the $y$ axis) and 
         $8\%$ (sides parallel to line $y=(x-1)/\sqrt{3}$) with respect to the 
         shown values.}             
         \blue{The $x$ and $y$ coordinates are in units of 
         $R_{\mathrm {ext}}$ and probability distribution in units of 
         $R_{\mathrm {ext}}^{-2}$. The distributions for $d=8$ nm are taken 
         as reference, and for the other cases they are magnified with 
         factors shown on the top left corner of each graph.}
         }
\label{fig:localization_non}         
\end{figure}

Many properties of polygonal quantum rings are very sensitive to the 
sample symmetry. Energy levels of non-symmetric triangular rings are only 
twofold (spin) degenerate. Still, probability distributions associated with the 
six lowest states, now arranged into three levels, are localized in corner 
areas. Contrary to the previous case they are not equally distributed 
between all of the corners but may be localized on single ones \cite{Sitek15}.  
In Fig.\ \ref{fig:energies_non} we show the energies of non-equilateral triangular 
samples for which the thickness of one side has been increased by $4\%$ 
and of a second one by $8\%$ with respect to the third side.
As can be seen in Fig.\ \ref{fig:energies_non}(a) the energy 
levels of the thicker sample resemble much more the ones of equilateral ring 
[Fig.\ \ref{fig:energies}(a)] than the energy structure of the thinner sample. 
This can be understood if one takes a look once again on the geometry of corner 
areas. If the ring sides have different thicknesses then the areas contained 
between internal and external boundaries differ from each other 
which leads to formation of three quantum wells 
\red{of different depth} within one ring.
In the case of narrower samples the ratio between 
corner areas is much higher than in the case of thicker rings, this results in 
substantially different quantum well 
\red{in each corner} and thus also energy spectrum and carrier 
localization do not resemble the ones of equilateral triangles. With increasing 
ring thickness the ratios between corner areas decrease and thus the 
difference between non-symmetric samples and equilateral rings becomes less 
pronounced and 
their properties similar. The positions of the ground state and exited levels as 
well as splittings between them change similarly to the case of equilateral 
triangles with only small shape dependent deviations 
[Fig.\ \ref{fig:energies_non}(b) and\ \ref{fig:energies_non}(c)], 
but the corresponding probability 
distributions differ considerably from the previously analyzed case. Irrespective 
of the aspect ratio the levels built of the six lowest states are separated 
from the higher ones by an energy gap and associated with corner-localized 
probability distributions while electrons occupying higher levels may be more 
easily found in the sides. Contrary to the symmetric samples, if the rings are 
sufficiently thin then, single localization peaks are formed in
both corner and side areas. 
\red{The deepest quantum well is formed in the corner for which the area between 
internal and external boundaries is the largest and thus the ground state electrons 
are localized in that corner [Fig.\ \ref{fig:localization_non}(a)]. The probability 
distribution associated with the second energy level forms a single peak around the 
medial corner area. Finally, the shallower quantum well is situated in the smallest 
corner whose ground state corresponds to the third energy level of the system and 
thus electrons excited to this level occupy the corner for which the area
between the boundaries is the smallest [Fig.\ \ref{fig:localization_non}(e)]. Also 
the three side areas differ from each other and the probability distribution  
associated with the first levels above the gap forms a single maximum in the largest
side area [Fig.\ \ref{fig:localization_non}(i)] while electrons excited to the two 
following levels occupy the two other sides, respectively. With increasing side 
thickness the 
ratio between the depth of the three quantum wells decreases which results in 
delocalized probability distributions and increasing number of localization peaks.}
For the geometry presented here probability distributions 
associated with the two lowest levels first form a single peak in the largest 
corner then two maxima at the ends of the 
thickest side and finally also a small maximum in the third corner
\red{[Fig.\ \ref{fig:localization_non}(a)-\ \ref{fig:localization_non}(d)]}.
Electrons 
excited to the third level practically occupy only the smallest corner but with 
very small probabilities they may also be found in the other corners of very 
thick samples 
\red{[Fig.\ \ref{fig:localization_non}(e)-\ \ref{fig:localization_non}(h)]}.
Localization patterns corresponding to the levels above the gap 
form one main peak in the widest side which is accompanied by two smaller peaks 
in the other sides of thick rings
\red{[Fig.\ \ref{fig:localization_non}(i)-\ \ref{fig:localization_non}(l)]}. As a 
result thicker non-symmetric rings resemble much more their equilateral 
counterparts and due to smaller differences between the effective quantum wells
it is much easier to control them externally, i.e., to achieve properties 
similar to equilateral counterparts as well as to thin non-symmetric rings.

 \subsection{\label{sec:absorption}Electromagnetic absorption}

\begin{figure}[t]\centering
    \includegraphics[width=0.48\textwidth,angle=0]{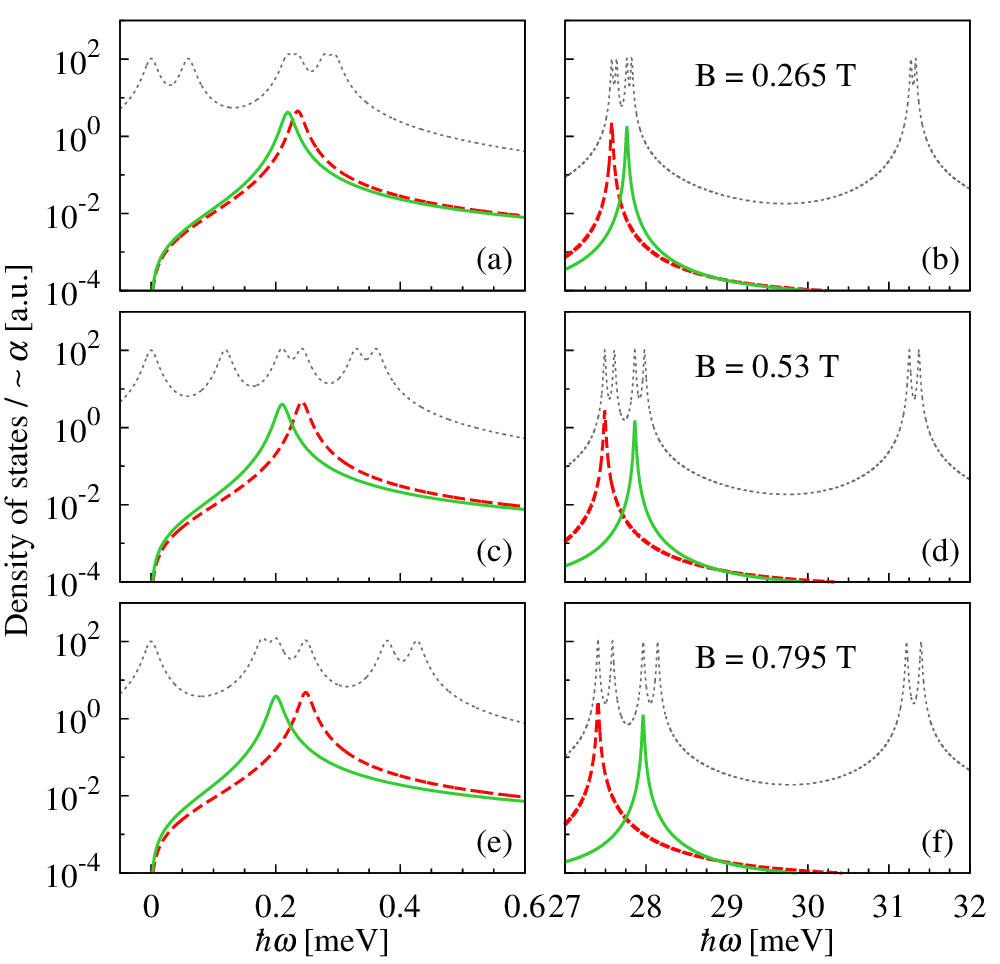}
\caption{Absorption coefficients associated with clockwise (green solid) 
         and counterclockwise (red dashed) polarization superimposed on the 
         density of states \blue{(gray dotted)} for a 12 nm thick equilateral ring 
         \blue{obtained}
         in the presence of external magnetic field. Values of the  field 
         shown in the right panels refer also to the left ones. 
         Figs. (a), (c) and (e) show transitions from the ground state to 
         other corner-localized states while Figs. (b), (d) and (f) 
         transitions to side-localized states. 
         For visibility we use a logarithmic scale for the absorption functions. 
         }    
\label{fig:abs_1}         
\end{figure}

We calculate the absorption coefficient using the formula 
\cite{Haug09,Chuang95,Hu00}
\begin{eqnarray}
\label{absorption_coeff}
 \alpha(\hbar\omega) = {\cal A}\hbar\omega\sum_{\mathrm{f}}|\langle f|\bm{\varepsilon}\cdot\bm{d}|i\rangle|^2
 \delta\left(\hbar\omega - \left(E_{\mathrm{f}}-E_{\mathrm{i}}\right)\right),
\end{eqnarray}
where ${\cal A}$ is a constant amplitude depending on the ring's properties,
$\bm{\varepsilon}=\left(1,\pm i\right)/\sqrt{2}$ the circular 
polarization of an electromagnetic field 
\red{(defined with respect to the ring's plane)},
$\bm{d}$ the dipole moment, and 
$E_{\mathrm{i,f}}$ the initial and final energies, respectively. 
\blue{ The dipole matrix elements depend on the eigenstates and thus are 
sensitive to the sample geometry,
\begin{eqnarray*}
\langle f|\bm{\varepsilon}\cdot\bm{d}|i\rangle = \frac{1}{\sqrt{2}}\sum_{q}
\Psi^{\dagger}\left(q,f\right)\Psi\left(q,i\right)r_k\left(\cos\phi_j\pm i \sin\phi_j\right).
\end{eqnarray*}
Here $q$ stands for the basis states ($|q\rangle\equiv |kj\sigma\rangle$)  
and $\Psi(q,a)$ are the amplitudes of the eigenvector 
$|a\rangle$ of the triangular ring in the $q$ basis, 
$|a\rangle=\sum_q\Psi\left(q,a\right)|q\rangle$. 
}

The $\delta$ (Dirac) function is approximated by a Lorentzian, 
\begin{eqnarray*}
 \delta\left(\hbar\omega -
 \left(E_{\mathrm{f}}-E_{\mathrm{i}}\right)\right)\approx
 \frac{\Gamma/2}{\left[\hbar\omega -
 \left(E_{\mathrm{f}}-E_{\mathrm{i}}\right)\right]^2
  + \left(\Gamma/2\right)^2} \ ,
    \label{sweight} 
\end{eqnarray*} 
where $\Gamma/2$ is a phenomenological broadening of the discrete energy 
levels of the samples and is fixed to $0.028$ meV. 
\blue{This value allows for a sufficient energy resolution for analyzing particular 
transitions. In practice this broadening may have complex physical origin, from 
all possible imperfections of the core-shell structure, but a proper modeling 
of it is beyond the scope of the present work.} 

\blue{ The absorption spectrum is governed by the intervals between the
energy levels, $E_f-E_i$, and by the dipole selection rules, incorporated
in Eq.\ (\ref{absorption_coeff}). Since the energy levels strongly depend
on side thickness or aspect ratio, the absorption of core-shell rings
may be established during the manufacturing process. It can be further
tuned with an external magnetic field perpendicular to the plane of the
polygon, which lifts both spin and orbital degeneracies. As a result the
four degenerate energy levels shown in Fig.\ \ref{fig:energies}(a) become
twelve non-degenerate levels with energies depending on the field strength.
In Fig.\ \ref{fig:abs_1} we compare energy intervals between
the ground state and the eleven excited states (around which maxima of
the gray dotted lines are formed) and absorption spectra (the green solid 
and red dashed lines) of a 12 nm symmetric quantum ring obtained for three 
different strengths of the magnetic field.  }

If a symmetric polygonal ring containing an electron in its ground 
state is exposed
to a clockwise polarized light then the electron can be excited to one of
the corner-localized or to one of the side-localized states. \blue{The
spin direction is conserved in the absence of spin-orbit coupling.}
Two different transitions occur in the presence of the 
counterclockwise polarized electromagnetic field, such that in total
the ground state electron may be excited to four different states, two
belonging to corner-localized domain \blue{[Figs. \ref{fig:abs_1}(a),
\ref{fig:abs_1}(c) and \ref{fig:abs_1}(e)]} and the other two to
side-localized one \blue{[Figs. \ref{fig:abs_1}(b), \ref{fig:abs_1}(d)
and  \ref{fig:abs_1}(f)]} \cite{Sitek15}.  As can be seen in Fig.\
\ref{fig:abs_1}, the positions of absorption maxima change with the
magnetic field on the single meV scale and \blue{ move apart with
increasing field strength. This can be a possibility to fine tuning the
absorbed energy. In the limiting case of vanishing magnetic field the
final excited levels merge in pairs  }
and form the second and third level, respectively, thus in the absence of 
the field one should expect two identical transitions from the ground state to 
those levels irrespective of polarization type. 

\begin{figure}[t]\centering
    \includegraphics[width=0.48\textwidth,angle=0]{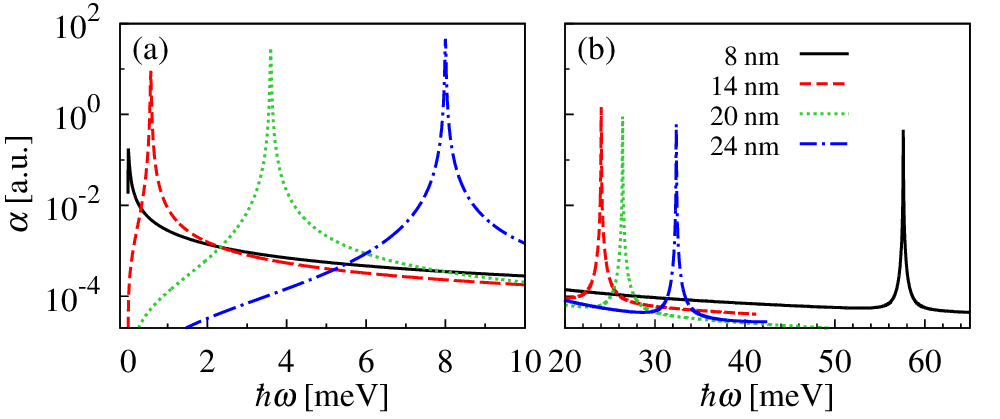}
\caption{Absorption coefficients associated with clockwise polarization and
         excitations from the ground state to other corner- (a) and 
         side-localized (b) states of equilateral triangular rings
         \blue{of different side thickness $d$. Values shown in 
         Fig. (b) refer to both panels.}
         $B=0.53$ T.         
         For visibility we use a logarithmic scale for the absorption functions.         
        }       
\label{fig:abs}         
\end{figure}

\blue{Excitations induced by both polarization types depend in a similar manner on 
side thickness and thus below we focus only on the effects induced by clockwise 
polarized 
electromagnetic field which excites the ground state electron to the third 
and ninth energy levels [the green solid lines in Fig.\ \ref{fig:abs_1}]. 
The first absorption peak is centered around the energy defined by the 
difference between the third and the first energy levels while the second 
maximum is formed at the energy equal to the splitting of the ninth and 
the ground level. The external magnetic field lifts the energy level degeneracy,
but all of the levels which in the absence of that field form one degenerate 
level change similarly with side thickness. In particular, the two 
lowest levels decay with increasing thickness like the ground state in the 
absence of the field [the red line in Fig.\ \ref{fig:energies}(b)] while the levels 
from seventh to tenth decrease like the third level of the degenerate 
system [the blue line in Fig.\ \ref{fig:energies}(b)]. As a result the positions 
of the two absorption peaks change with side thickness similarly to the 
splittings of the relevant degenerate levels [Fig.\ \ref{fig:energies}(c)].}

\blue{The energy difference between the ground state and the third excited level 
increases with side thickness and thus absorption maxima associated with 
this transition [Fig.\ \ref{fig:abs}(a)] move to higher energies.
The energy interval which defines the position of the second peak
depends on side thickness like the splitting between the third and ground 
levels of degenerate systems [blue dotted line in Fig.\ \ref{fig:energies}(c)].
Corner- and side-localized states of thin triangular rings are separated from 
each other by a large energy gap, as a consequence the splitting between the 
ground state and the first optically accessible side-localized state is also
considerably large and thus the second absorption peak occurs at 
high energies [the black solid line in Fig.\ \ref{fig:abs}(b)]. With increasing
side thickness the relevant energy splitting decreases, as a result the 
second absorption peak initially shifts to lower energies. The situation 
changes when the aspect ratio corresponding to the minimal difference 
between the initial and final levels is reached. From this
point the absorption maximum moves back to higher energies, because 
the energy difference between the ground state and the ninth level
increases 
with side thickness for sufficiently thick samples, Fig.\ \ref{fig:abs}(b). 
}

This means that it is possible to achieve two rings for which aspect ratios 
differ considerably but the samples still have one transition induced by photons
of the same frequency. 
Irrespective of the sample shape the first transition requires absorption of 
electromagnetic waves from the THz region. On the contrary, excitations to 
the side-localized states show much stronger dependence on the sample geometry 
and may be produced by THz and infrared electromagnetic 
waves as well. This shows that the absorption spectrum of 
triangular quantum rings may cover wide range of wavelengths and may be 
engineered during the manufacturing process through side thickness and then, 
more precisely, by external magnetic or electric 
fields.

\begin{figure}[t]\centering
    \includegraphics[width=0.48\textwidth,angle=0]{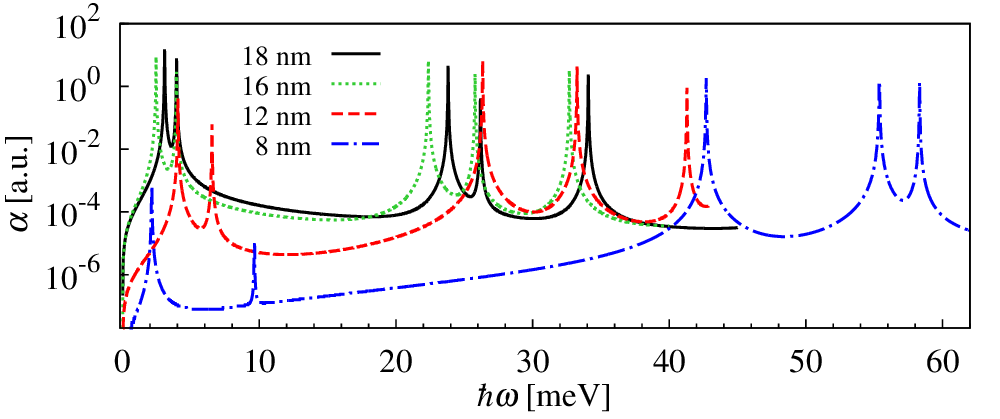}
\caption{Absorption coefficients associated with clockwise polarization 
         and excitations from the ground state to the first five excited \blue{levels}
         of non-symmetric samples, defined in Figs.\ \ref{fig:energies_non}
         and \ \ref{fig:localization_non}. The numbers shown in the figure 
         refer to the thickness of the narrowest sides, $B=0$.
         For visibility we use a logarithmic scale for the absorption functions.         
        }       
\label{fig:abs_non}         
\end{figure}

In the case of equilateral rings (Fig.\ \ref{fig:abs}) the absorption 
coefficients associated with both transitions differ up to one order of 
magnitude throughout the whole range of analyzed side thicknesses. This 
situation does not hold for non-symmetric samples. The absorption coefficients 
(Eq.\ \ref{absorption_coeff}) depend on energy gaps
and on dipole moments. The former ones 
change similarly in both cases [Figs.\ \ref{fig:energies}(c) and 
\ref{fig:energies_non}(c)], but wave functions 
[localization shown in Figs.\ \ref{fig:localization} 
and\ \ref{fig:localization_non}]
and thus dipole moments are 
very sensitive to the sample symmetry. As a result the absorption spectrum 
of non-symmetric rings (Fig.\ \ref{fig:abs_non}) differs considerably from 
the one of equilateral samples (Fig.\ \ref{fig:abs}). 
The first consequence of the broken wave function symmetry is the 
increased number of optically 
induced transitions. In principle all spin-allowed excitations occur, 
i.e., the ground state electron may be excited to the two higher levels 
associated with localization in corner areas and to the three levels 
corresponding to mostly side-localized probability distributions
\blue{(Fig.\ \ref{fig:abs_non})}.
\blue{Moreover this occurs} in the presence of both polarization types. 
\blue{The positions of the absorption peaks are determined by
the energy intervals between the ground state and the five excited 
levels [Fig.\ \ref{fig:energies_non}(a)] and change with side thickness. 
For example, the position of the first absorption peak is determined by the 
energy separation between the second energy level and the ground state 
while the third maximum occurs for energy equal to the difference between 
the third and first levels. The energies at which those two absorption peaks 
built up depend on side thickness like the green and blue lines in 
Fig.\ \ref{fig:energies_non}(c), respectively}
Thin samples absorb much stronger 
photons of higher frequencies which allow for excitation of the ground 
state electrons (localized around
a single corner) to states associated with 
probability distributions forming single maxima in the middle of each side. 
For such samples the absorption in the corner-state domain in negligibly 
small. The absorption coefficients corresponding to transitions to other 
corner-localized states constantly increase with ring thickness
such that for thick samples they exceed the ones occurring at higher 
frequencies. The strong dependence of the absorption coefficient on aspect 
ratio suggests that samples with different optical properties may be
relatively easily obtained by slightly changing the deposition time during
the fabrication of the core-shell structure.

\section{\label{sec:conclusions}Conclusions}

\blue{
We studied how sample geometry details affect energy levels and the corresponding
probability distributions of a single electron confined in a triangular quantum 
ring. We also showed how those features determine optical absorption of circularly 
polarized electromagnetic field. The ground state of triangular samples restricted 
internally and externally by regular polygons is twofold degenerate and is followed 
by alternating sequence of pairs of four- and twofold degenerate levels. The energy 
degeneracy depends only on the ring symmetry and the fourfold degenerate levels may
be split into pairs of spin degenerate eigenvalues if the symmetry is broken by, e.g.,
different side thicknesses. The spin degeneracy is lifted when the sample is 
immersed in external magnetic field. The splittings between consecutive levels strongly 
depend on aspect ratio. If the external radius is assumed to be constant then, the 
separations between the ground state and all other states associated with corner-localized
probability distributions increases with side thicknesses. The main energy gap which 
separates corner- from side-localized states decreases with the side thickness until
the lowest energy gap,
occurring for the aspect ratio of about $0.34$,
is reached and then the gap increases slightly again. Contrary to other polygonal rings, 
in the case of triangular samples this gap is always considerably larger than the 
energy splittings in the corner state domain. The position of the ground state, 
followed by the excited levels, is shifted to lower energies with increasing side 
thickness. Electron localization is also very sensitive to the sample shape. Low-energy 
electrons confined in thin rings occupy only very small corner areas, with increasing side 
thickness the corresponding localization peaks penetrate into side areas and overlap.
The probability distributions associated with energy levels above the main gap change 
in the opposite way, i.e., wide peaks spread along side areas become sharper with 
increasing aspect ratio and for sufficiently thick rings are accompanied by lower 
maxima formed around corner areas. }

\blue{
The energy levels and corresponding wave functions govern absorption spectrum of the
systems. The positions of absorption peaks are determined by energy splittings between 
pairs of particular levels, while their amplitudes depend on dipole moments
and thus on wave functions. As a result the energy of absorbed electromagnetic 
field depends on sample shape. Since the corner- and side-localized states are 
always separated by a considerable energy gap, excitations of an electron from 
the ground state to other corner-localized states and side-localized states 
occurs in the presence of photons associated with wavelengths belonging to 
different electromagnetic domains.  
Thus, by selecting the appropriate ring thickness and shape, one can engineer 
the absorption in the THz and in the infrared domain.}

\blue{
For simplicity of description and discussion we keep
the external radius constant and change only the side thickness. The way
how the energy levels change does not depend on the ring diameter but
only on the aspect ratio between the side thickness and the external
radius. Nonetheless energy levels of nanoscale systems do depend on
sample sizes, and thus the results shown in Fig.\ \ref{fig:energies}
calculated for different external radii would look the same, but the
axes would be scaled differently.}

\red{
Throughout this work the electron-electron Coulomb interaction was neglected. 
Strictly speaking the results presented correspond to transitions of 
a single electron in the triangular ring. As long as the Coulomb interaction is small, 
possibly due to a large dielectric constant of the material, its main effect 
is a renormalization of the energy spectrum. However, because of the small 
energy dispersion of the corner states, non-trivial Coulomb effects could 
be expected.
}

\vspace{0.7 cm}
\noindent
\textbf{\Large{Acknowledgement}}
\vspace{0.2 cm}

\noindent
This work was mainly supported by the Research Fund
of the University of Iceland. Additional funding from the Nordic network 
NANOCONTROL, project No.: P-13053, and COST Action MP1204 'Tera-MIR 
Radiation: Materials, Generation, Detection and Applications' is acknowledged.


\begin{thebibliography}{10}
\expandafter\ifx\csname url\endcsname\relax
  \def\url#1{{\tt #1}}\fi
\expandafter\ifx\csname urlprefix\endcsname\relax\def\urlprefix{URL }\fi
\providecommand{\eprint}[2][]{\url{#2}}

\bibitem{Shi15}
Shi T, Jackson H~E, Smith L~M, Jiang N, Gao Q, Tan H~H, Jagadish C, Zheng C and
  Etheridge J 2015 {\em Nano Letters\/} {\bf 15} 1876--1882 pMID: 25714336
  (\textit{Preprint} \eprint{http://dx.doi.org/10.1021/nl5046878})
  \urlprefix\url{http://dx.doi.org/10.1021/nl5046878}

\bibitem{Krogstrup13}
Krogstrup P, Jorgensen H~I, Heiss M, Demichel O, Holm J~V, Aagesen M, Nygard J
  and Fontcuberta~i Morral A 2013 {\em Nature Photonics\/} {\bf 7} 306--310
  \urlprefix\url{http://dx.doi.org/10.1038/nphoton.2013.32}

\bibitem{Tang11}
Tang J, Huo Z, Brittman S, Gao H and Yang P 2011 {\em Nature Nanotechnology\/}
  {\bf 6} 568--572 \urlprefix\url{http://dx.doi.org/10.1038/nnano.2011.139}

\bibitem{Kim15}
Kim S~K, Zhang X, Hill D~J, Song K~D, Park J~S, Park H~G and Cahoon J~F 2015
  {\em Nano Letters\/} {\bf 15} 753--758 pMID: 25546325 (\textit{Preprint}
  \eprint{http://dx.doi.org/10.1021/nl504462e})
  \urlprefix\url{http://dx.doi.org/10.1021/nl504462e}

\bibitem{Xiang06}
Xiang J, Lu W, Hu Y, Wu Y, Yan H and Lieber C~M 2006 {\em Nature\/} {\bf 441}
  489--493

\bibitem{Nguyen14}
Nguyen B~M, Taur Y, Picraux S~T and Dayeh S~A 2014 {\em Nano Letters\/} {\bf
  14} 585--591 (\textit{Preprint}
  \eprint{http://pubs.acs.org/doi/pdf/10.1021/nl4037559})
  \urlprefix\url{http://pubs.acs.org/doi/abs/10.1021/nl4037559}

\bibitem{Saxena13}
Saxena D, Mokkapati S, Parkinson P, Jiang N, Gao Q, Tan H~H and Jagadish C 2013
  {\em Nat Photon\/} {\bf 7} 963--968
  \urlprefix\url{http://dx.doi.org/10.1038/nphoton.2013.303}

\bibitem{Ho15}
Ho J, Tatebayashi J, Sergent S, Fong C~F, Iwamoto S and Arakawa Y 2015 {\em ACS
  Photonics\/} {\bf 2} 165--171 (\textit{Preprint}
  \eprint{http://dx.doi.org/10.1021/ph5003945})
  \urlprefix\url{http://dx.doi.org/10.1021/ph5003945}

\bibitem{Thierry12}
Thierry R, Perillat-Merceroz G, Jouneau P~H, Ferret P and Feuillet G 2012 {\em
  Nanotechnology\/} {\bf 23} 085705
  \urlprefix\url{http://stacks.iop.org/0957-4484/23/i=8/a=085705}

\bibitem{Ibanes13}
John~Ibanes J, Herminia~Balgos M, Jaculbia R, Salvador A, Somintac A, Estacio
  E, Que C~T, Tsuzuki S, Yamamoto K and Tani M 2013 {\em Applied Physics
  Letters\/} {\bf 102} 063101
  \urlprefix\url{http://scitation.aip.org/content/aip/journal/apl/102/6/10.1063/1.4791570}

\bibitem{Peng15}
Peng K, Parkinson P, Fu L, Gao Q, Jiang N, Guo Y~N, Wang F, Joyce H~J, Boland
  J~L, Tan H~H, Jagadish C and Johnston M~B 2015 {\em Nano Letters\/} {\bf 15}
  206--210 pMID: 25490548 (\textit{Preprint}
  \eprint{http://dx.doi.org/10.1021/nl5033843})
  \urlprefix\url{http://dx.doi.org/10.1021/nl5033843}

\bibitem{Chuang95}
Chuang S~L 1995 {\em Physics of Optoelectronic Devices\/} (New York: John Wiley
  and Sons, Inc.)

\bibitem{Pistol08}
Pistol M~E and Pryor C~E 2008 {\em Phys. Rev. B\/} {\bf 78}(11) 115319
  \urlprefix\url{http://link.aps.org/doi/10.1103/PhysRevB.78.115319}

\bibitem{Wong11}
Wong B~M, L{\'e}onard F, Li Q and Wang G~T 2011 {\em Nano Letters\/} {\bf 11}
  3074--3079 pMID: 21696178 (\textit{Preprint}
  \eprint{http://dx.doi.org/10.1021/nl200981x})
  \urlprefix\url{http://dx.doi.org/10.1021/nl200981x}

\bibitem{Blomers13}
Bl{\"o}mers C, Rieger T, Zellekens P, Haas F, Lepsa M~I, Hardtdegen H, G{\"u}l
  {\"O}, Demarina N, Gr{\"u}tzmacher D, L{\"u}th H and Sch{\"a}pers T 2013 {\em
  Nanotechnology\/} {\bf 24} 035203
  \urlprefix\url{http://stacks.iop.org/0957-4484/24/i=3/a=035203}

\bibitem{Rieger12}
Rieger T, Luysberg M, Sch{\"a}pers T, Gr{\"u}tzmacher D and Lepsa M~I 2012 {\em
  Nano Letters\/} {\bf 12} 5559--5564 pMID: 23030380 (\textit{Preprint}
  \eprint{http://dx.doi.org/10.1021/nl302502b})
  \urlprefix\url{http://dx.doi.org/10.1021/nl302502b}

\bibitem{Haas13}
Haas F, Sladek K, Winden A, von~der Ahe M, Weirich T~E, Rieger T, L{\"u}th H,
  Gr{\"u}tzmacher D, Sch{\"a}pers T and Hardtdegen H 2013 {\em
  Nanotechnology\/} {\bf 24} 085603
  \urlprefix\url{http://stacks.iop.org/0957-4484/24/i=8/a=085603}

\bibitem{Sprung}
Sprung D~W~L, Wu H and Martorell J 1992 {\em J. Appl. Phys.\/} {\bf 71}(1)
  515--517

\bibitem{Ballester12}
Ballester A, Planelles J and Bertoni A 2012 {\em Journal of Applied Physics\/}
  {\bf 112} 104317
  \urlprefix\url{http://scitation.aip.org/content/aip/journal/jap/112/10/10.1063/1.4766444}

\bibitem{Sitek15}
Sitek A, Serra L, Gudmundsson V and Manolescu A 2015 {\em Phys. Rev. B\/} {\bf
  91}(23) 235429
  \urlprefix\url{http://link.aps.org/doi/10.1103/PhysRevB.91.235429}

\bibitem{Estarellas15}
Estarellas C and Serra L 2015 {\em Superlattice Microstruct\/} {\bf 83} 184

\bibitem{Daday11}
Daday C, Manolescu A, Marinescu D~C and Gudmundsson V 2011 {\em Phys. Rev. B\/}
  {\bf 84}(11) 115311
  \urlprefix\url{http://link.aps.org/doi/10.1103/PhysRevB.84.115311}

\bibitem{Groth14}
Groth C~W, Wimmer M, Akhmerov A~R and Waintal X 2014 {\em New Journal of
  Physics\/} {\bf 16} 063065
  \urlprefix\url{http://stacks.iop.org/1367-2630/16/i=6/a=063065}

\bibitem{Haug09}
Haug H and Koch S~W 2009 {\em Quantum Theory of the Optical and Electronic
  Properties of Semiconductors\/} 5th ed (Singapore: World Scientific)

\bibitem{Hu00}
Hu H, Zhu J~L and Xiong J~J 2000 {\em Phys. Rev. B\/} {\bf 62}(24) 16777--16783
  \urlprefix\url{http://link.aps.org/doi/10.1103/PhysRevB.62.16777}

\end{thebibliography}
\providecommand{\newblock}{}

\end{document}